\documentclass[aps,showpacs,prl,twocolumn,twoside,groupedaddress]{revtex4}
\usepackage{graphicx}
\usepackage{epsfig}
\usepackage{epsf}
\usepackage{amssymb}
\usepackage{epstopdf}
\usepackage{amsmath}

\begin{document}

\title{Protecting quantum entanglement from amplitude damping}

\author{Zeyang Liao$^{1}$, M. Al-Amri$^{1, 2}$, and M. Suhail Zubairy$^{1}$}

\affiliation{$^1$Institute for Quantum Science and Engineering (IQSE) and Department of Physics and Astronomy, Texas A$\&$M University, College Station, TX 77843-4242, USA\\
$^2$The National Center for Mathematics and Physics, KACST, P.O.Box 6086, Riyadh 11442, Saudi Arabia}

\begin{abstract}
Quantum entanglement is a critical resource for quantum information and quantum computation. However, entanglement of a quantum system is subjected to change due to the interaction with the environment. One typical result of the interaction is the amplitude damping that usually results in the reduction of the entanglement. Here we propose a protocol to protect quantum entanglement from the amplitude damping by applying Hadamard and CNOT gates. As opposed to some recently studied methods, the scheme presented here does not require weak measurement in the reversal process, leading to a faster recovery of entanglement. We propose a possible experimental implementation based on linear optical system. 
\end{abstract}

\pacs{03.67.Bg, 03.65.Yz, 42.50.Ex, 03.65.Ta, 03.67.Pp} \maketitle

\section{I. introduction}

A quantum state is subjected to decoherence due to the interaction with the environment. The quantum error correction codes can suppress the decoherence by encoding the logical qubit in multiple physical qubits and performing sufficient measurements and correction operations \cite{Shor1995, Steane77, Shor1996, Steane452}. Another strategy is to rely on the so-called decoherence-free subspace which requires the interaction Hamiltonian to have some appropriate symmetry  \cite{Lidar, Kwiat}.  quantum Zeno effect\cite{Facchi, Maniscalco} and dynamical decoupling \cite{Viola} have also been discussed to protect the quantum state. 

Amplitude damping is one important type of decoherence which is related to many practical qubit systems \cite{Weisskopf}. For example, it can happen to a photon qubit in a leaky cavity, or atomic qubit subjected to spontaneous decay, or a superconduction qubit with zero-temperature energy relaxation. Recently, it is shown that weak measurement together with bit flip can recover a quantum state from amplitude damping \cite{Korotkov2006, Katz, Kim2009, Sun1}. Another way to restore a qubit state in a weak measurement is using Hadamard and CNOT gates with auxiliary qubits \cite{Amri}. In this way, no weak measurement is required in the reversal process, and hence the reversal time can be shorter.

Quantum entanglement, which is a critical resource of the quantum information and quantum computation, also decreases due to the amplitude damping.  Sun {\it{et. al}} first showed that weak measurement together with bit flip can also protect the quantum entanglement \cite {Sun2}. The decoherence can be largely suppressed by uncollapsing the quantum state toward ground state before the amplitude damping \cite{Korotkov2010}. These ideas were recently implemented in a proof-of-principle experiment \cite{Kim}. 

In this paper, we show that, following the approach presented in \cite{Amri}, an arbitrary two-qubit pure state under amplitude damping in a weak measurement can also be probabilistically recovered using Hadamard and CNOT gates with auxiliary qubits. Furthermore, even without weak measurement, quantum entanglement of a two-qubit system under amplitude damping can also be partially protected using our scheme.  We also propose a proof-of-principle experiment for this scheme based on linear optical system. Besides, we can extend our scheme to suppress the decoherence even better by uncollapsing the quantum state of the system toward the ground state.

This paper is organized as follows: In Sec. II we briefly introduce the weak measurement and amplitude damping. In Sec. III we show that a pure two-qubit state that has undergone a weak measurement can be recovered using Hadamard and CNOT gates with auxiliary qubits. In Sec. IV we show that quantum entanglement between two qubits that has undergone amplitude damping can be partially recovered using the same procedure as presented in Sec. III. In Sec. V we propose a linear optical experiment to implement our scheme. In Sec. IV, we present an extended scheme to protect entanglement with better efficiency. In Sec. VII we discuss the fidelity of the recovered entangled state. Finally we summarize the result.  

\section{II. Weak measurement and amplitude damping}

As opposed to a typical Von Neumann quantum measurement, complete collapse to an eigenstate does not occur in a weak measurement \cite{Korotkov1999}. An example of the weak measurement is the leakage of the field inside a cavity. Suppose that the quantum state of a field in a cavity is a supperposition of zero and one photon states, i.e., $|\psi \rangle=\alpha |0 \rangle +\beta |1 \rangle$ with $|\alpha |^{2}+|\beta|^{2}=1$. Let us assume that an ideal detector is placed outside the cavity. If the detector registers a click, the quantum state of the field collapses to $|1\rangle$. However, if no click happens, the quantum state does not collapse but is reduced to $|\psi(\tau)\rangle=(\alpha |0\rangle+\beta \exp(-\Gamma \tau)|1\rangle)/\sqrt{|\alpha|^{2}+|\beta|^{2}\exp(-2\Gamma \tau)}$ where $\Gamma$ is the cavity decay rate.  The amplitude of the one photon state is damped. 

More generally, an amplitude damping of a single qubit can be described by the following mapping \cite{Weisskopf}: 
\begin{eqnarray}
|0\rangle _{S}|0\rangle _{E} & \rightarrow & |0\rangle _{S}|0\rangle _{E}, \\
|1\rangle _{S}|0\rangle _{E} & \rightarrow & \sqrt{q}|1\rangle _{S}|0\rangle _{E}+\sqrt{p}|0\rangle _{S}|1\rangle _{E}, 
\end{eqnarray}
where $p\in [0,1]$ is the possibility of decaying of the excited state, $q=1-p$ and S (E) denotes the system (environment). Within the Weisskopf-Wigner approximation, the probability of finding the atom in the excited state decreases exponentially with time and we have $\sqrt{q}=e^{-\Gamma t}$. 

In a weak measurement, if a detector gets a null-result, we have the following mapping:
\begin{eqnarray}
|0\rangle _{S}|0\rangle _{E} & \rightarrow & |0\rangle _{S}|0\rangle _{E}, \\ 
|1\rangle _{S}|0\rangle _{E} & \rightarrow & \sqrt{q}|1\rangle _{S}|0\rangle _{E}. 
\end{eqnarray}

\section{III. Two-qubit state recovery in a weak measurment}

In this section, we consider an arbitrary two-qubit pure state which is given by
\begin{equation}
|\psi \rangle_{in}=\alpha |00\rangle _{S} + \beta |01\rangle _{S} + \gamma |10\rangle _{S}+\delta |11\rangle _{S}.
\end{equation}

When this state undergoes amplitude damping, the amplitudes are modified. We consider two situations: In this section, we consider the case when we get a null-result for the weak measurement and the system evolves according to the mapping Eqs. (3) and (4). In the next section we will consider the case when the two-qubit state undergoes a general amplitude damping which is described by Eqs. (1) and (2).

When we get a null result in the amlitude damping, the system evolves according to the mappings in Eqs. (3) and (4) to
\begin{equation}
|\psi \rangle_{d}=\frac{1}{N_{d}}(\alpha |00\rangle _{S}+ \beta \sqrt{q}|01\rangle _{S}+ \gamma \sqrt{q}|10\rangle _{S}+\delta q|11\rangle _{S})
\end{equation}
where $N_{d}=\sqrt{|\alpha|^{2}+q(|\beta|^{2}+|\gamma|^{2})+q^{2}|\delta|^{2}}$ is the normalization factor. 

\begin{figure}
\includegraphics[width=0.95\columnwidth]{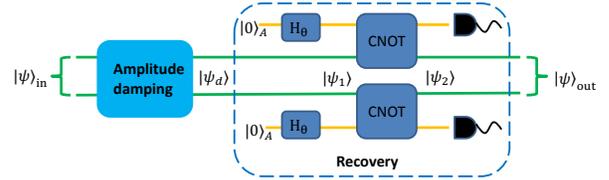}
\caption{(Color online) Circuit diagram for the reversal of the weak measurement and the quantum entanglement protection using Hadamard and CNOT gates. } 
\end{figure}

To recover the damped quantum state back to the initial quantum state, we use the circuit diagram shown in Fig. 1. Two auxiliary qubits are needed in this scheme. Initially, these two ancillas are both in the $|0\rangle $ state. First we apply a Hadamard gate with angle $\theta$ for each ancilla
\begin{equation}
 H_{\theta}=\left( \begin{array}{cc}
\cos\theta & -\sin\theta \\
\sin\theta & \cos\theta  \end{array} \right).
\end{equation}
The combined system is given by
\begin{eqnarray}
|\psi_{1} \rangle &=&\frac{1}{N_{d}}[\alpha |00\rangle _{S}+ \beta \sqrt{q} |01\rangle _{S}+ \gamma \sqrt{q}|10\rangle_{S} +\delta q|11\rangle_{S} ] \nonumber \\ &\otimes & (\cos\theta |0\rangle _{A}+\sin\theta |0\rangle _{A})(\cos\theta |0\rangle _{A}+\sin\theta |0\rangle _{A}). 
\end{eqnarray}

Then two CNOT gates are separately applied to each pair of the system qubit and the ancilla qubit. The system qubits are the controlled qubits while the ancilla qubits are the target qubits.
If $\theta$ is chosen to be $\tan^{-1}(1/\sqrt{q})$ or $\tan^{-1}[\exp(\Gamma t)]$, the combined state becomes (See appendix A)
\begin{widetext}
\begin{eqnarray}
|\psi_{2}\rangle&=& \frac{q}{N_{d}(1+q)} (\alpha |00\rangle _{S}+ \beta |01\rangle _{S}+ \gamma |10\rangle _{S}+\delta |11\rangle _{S}) |00\rangle _{A}
+  \frac{\sqrt{q}}{N_{d}(1+q)}[\alpha |00\rangle _{S}+ \beta q |01\rangle _{S}+ \gamma |10\rangle _{S}+\delta q|11\rangle _{S}] |01\rangle _{A} \nonumber \\ 
&+ &\frac{\sqrt{q}}{N_{d}(1+q)}[\alpha |00\rangle _{S}+ \beta |01\rangle _{S}+  q(\gamma |10\rangle _{S}+\delta |11\rangle _{S})]|10\rangle _{A}+ \frac{1}{N_{d}(1+q)}[\alpha |00\rangle _{S}+ q(\beta  |01\rangle _{S}+ \gamma |10\rangle _{S})+\delta q^2 |11\rangle _{S}] |11\rangle _{A} \nonumber \\ 
\end{eqnarray}
\end{widetext} 
After the CNOT gates, we make a measurement on the ancilla qubits. From Eq. (9), we can see that if we get $|00\rangle $ result, the state of the system recovers back to the initial state exactly. The success probability is $P_{00}(q)=[q/N_{d}(1+q)]^{2}$ which decreases with the decaying probability (see Fig. 2). 

If we get $|01\rangle $ or $|10\rangle $ for the ancilla qubit, we just repeat the same procedure on one qubit and with $\theta =\tan^{-1} (1/q)$. For example, if we get $|01\rangle $ for the ancilla qubit, the quantum state of the system is $|\psi\rangle_{out}=\alpha |00\rangle + \beta q |01\rangle + \gamma |10\rangle +\delta q|11\rangle $. This state can be interpreted as if the first qubit has not decayed while the second qubit has a decay rate $2\Gamma$.  In this case we add another ancilla for the second qubit. By applying a Hadamard gate with $\theta =\tan^{-1} (1/q)$ for the ancilla and CNOT gate for the ancilla and the second qubit, we obtain the following state
\begin{eqnarray}
 |\psi_{2}^{'}\rangle &=& \frac{q}{\sqrt{1+q^{2}}} (\alpha|00 \rangle _{S}+\beta |01 \rangle _{S}+\gamma |10\rangle _{S}+\delta |11 \rangle _{S})|0\rangle _{A} \nonumber \\ &+& \frac{1}{\sqrt{1+q^{2}}} (\alpha|00 \rangle _{S}+\beta q^{2} |01 \rangle _{S}+\gamma |10\rangle _{S}+\delta q^{2} |11 \rangle _{S})|1\rangle _{A} \nonumber \\
\end{eqnarray} 
We can see that the state of the system can also be recovered if the state of ancilla qubit is measured to be $|0\rangle$. The success probability in this case is given by $P_{010}(q)=q^{2}/(1+q^{2})$. If we get $|1\rangle$ on the ancilla qubit, the effective decay rate is double and we can repeat the same procedure but with $\theta=\tan^{-1}(1/q^{2})$. Repeating the same procedure again and again we can increase the probability to recover the quantum state.

If we get $|11\rangle $ result, the state of the system is given by the last term in Eq. (9). Comparing this term with Eq. (6) we find that the only difference is that the damping coefficient $\sqrt{q}$ is replaced by $q$ which means that the decay rate is double. Therefore, by repeating the same procedure but with $\theta = \tan^{-1}(1/q)$, we can still recover the initial state with a reduced probability.   

The probability to recover the quantum state versus the the damping rate for different repeat times is shown in Fig. 2 (see appendix A for the calculations). From the figure we can see that the success probability decreases as the damping rate increases. When the quantum state is completely damped ($p=1$), we can never recover the state back because the information of the state has been lost. We can also see that the success probability can be significantly increased for the first few repeating times. When we increase the repeating time we can approach the success probability of double weak measurement ($q^{2}$) which is shown by the dashed line. In practice we can repeat about three times and we can have significant success probability. After three times the success probability increases by only very small amounts especially when the damping rate is large.

\begin{figure}
\includegraphics[width=0.95\columnwidth]{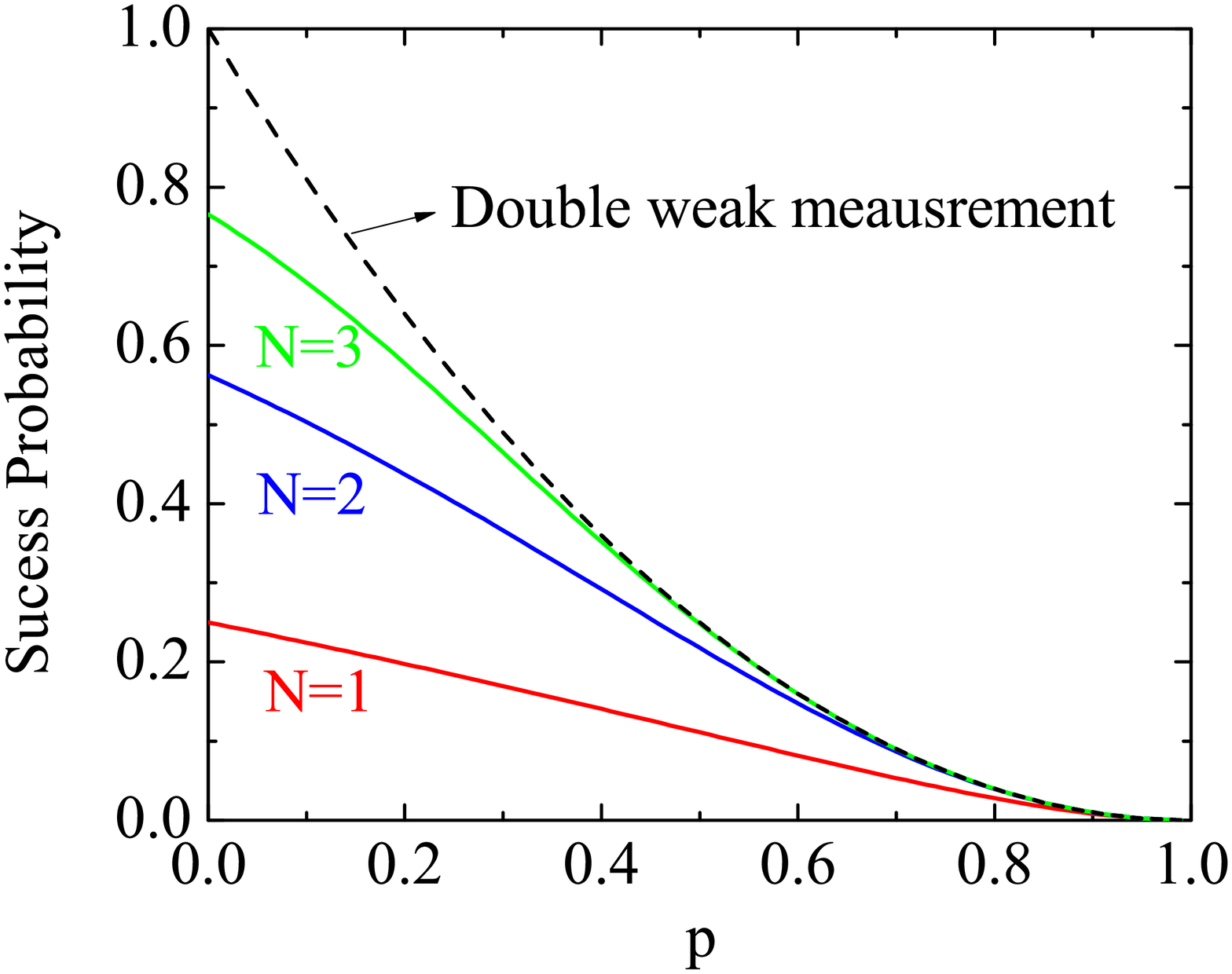}
\caption{(Color online) The success probability versus the the damping rate for different repeat times. The dashed line is the success probability for double weak measurement. } 
\end{figure}

\section{IV. two-qubit quantum entanglement protection from amplitude damping}

In the previous section, we showed that a two-qubit quantum state in a weak measurement can be recovered probabilistically using the procedure shown in Fig. 1. In this section we show that using similar procedure we can protect two-qubit entanglement when the two-qubit system undergoes a general amplitude damping which is described by the mappings Eqs. (1) and (2). 

As discussed in the previous section, a general two-qubit pure state is given by $|\psi\rangle_{in}=\alpha |00\rangle + \beta |01\rangle + \gamma |10\rangle +\delta |11\rangle $. The concurrence of this state is $C_{i}=\max\{0, 2|\alpha \delta - \beta \gamma|\}$. We assume that the environment is in the ground state $|0\rangle$. The evolution of the combined system with amplitude damping is given by
\begin{eqnarray}
|\psi_{d}\rangle &=& \alpha |00\rangle _{S}|00\rangle _{E}+ \beta \sqrt{q}|01\rangle _{S}|00\rangle _{E}+\beta \sqrt{p}|00\rangle _{S}|01\rangle _{E} \nonumber \\ & + &\gamma \sqrt{q}|10\rangle _{S}|00\rangle _{E}  +\gamma \sqrt{p}|00\rangle _{S}|10\rangle _{E}+\delta q|11\rangle _{S}|00\rangle _{E} \nonumber \\
& + &\delta \sqrt{qp}|10\rangle _{S}|01\rangle _{E}+\delta \sqrt{qp}|01\rangle _{S}|10\rangle _{E}+\delta p|00\rangle _{S}|11\rangle _{E}. \nonumber \\
\end{eqnarray}
On tracing out the environment we obtain the density matrix for the system and from which we can get the damped concurrence \cite{Tahira}
\begin{equation} 
C_{d}(p)=\max\{0,2q(|\alpha\delta-\beta\gamma|-p|\delta|^2)\}.
\end{equation}
This concurrence is less than the initial value and decreases with an increasing $p$. This indicates that the entanglement of the system decreases due to the amplitude damping. 

To protect the entanglement, we add two ancilla qubits which are initially in the $|00\rangle $ state and follow the same procedure as in the previous section. We measure the final state of the ancilla qubit. There are four possible outcomes. If the result is $|00 \rangle$, we obtain the density matrix of the system after tracing out the environment to be
\begin{widetext}
\begin{equation}
\rho _{f}=\frac{1}{N_{2}} \left( \begin{array}{cccc}
|\alpha|^{2}+p|\beta|^{2}+p|\gamma|^{2}+p^{2}|\delta|^{2} & \alpha \beta^{*}+p\gamma\delta^{*} & \alpha \gamma^{*}+p\beta\delta^{*} &  \alpha \delta^{*} \\
\alpha^{*}\beta + \gamma^{*} \delta p & |\beta|^{2}+|\delta|^{2} p & \beta \gamma^{*}  & \beta \delta^{*} \\
 \alpha ^{*} \gamma + p\beta ^{*}\delta & \beta ^{*} \gamma & |\gamma|^{2}+ p|\delta|^{2} & \gamma \delta ^{*} \\
\alpha ^{*} \delta & \beta^{*} \delta & \gamma^{*} \delta &  |\delta|^{2} 
\end{array} \right)
\end{equation}
\end{widetext}
where $N_{2}=1+p(|\beta|^{2}+|\gamma|^{2}+2|\delta|^{2})+p^{2}|\delta|^{2}$. We note that this result is identical to the result as obtained via weak measurement reversal \cite{Sun2}. The probability to get this result is $[q/N_{2}(1+q)]^{2}$.  The concurrence of the final state is given by
\begin{equation}
C_{r}(p)=\max \{ 0,\frac{2(|\alpha\delta-\beta\gamma|-p|\delta|^{2})}{1+p(1+|\delta|^{2}-|\alpha|^{2})+p^{2}|\delta|^{2}}\}.
\end{equation}

\begin{figure*}
\includegraphics[width=0.9\columnwidth]{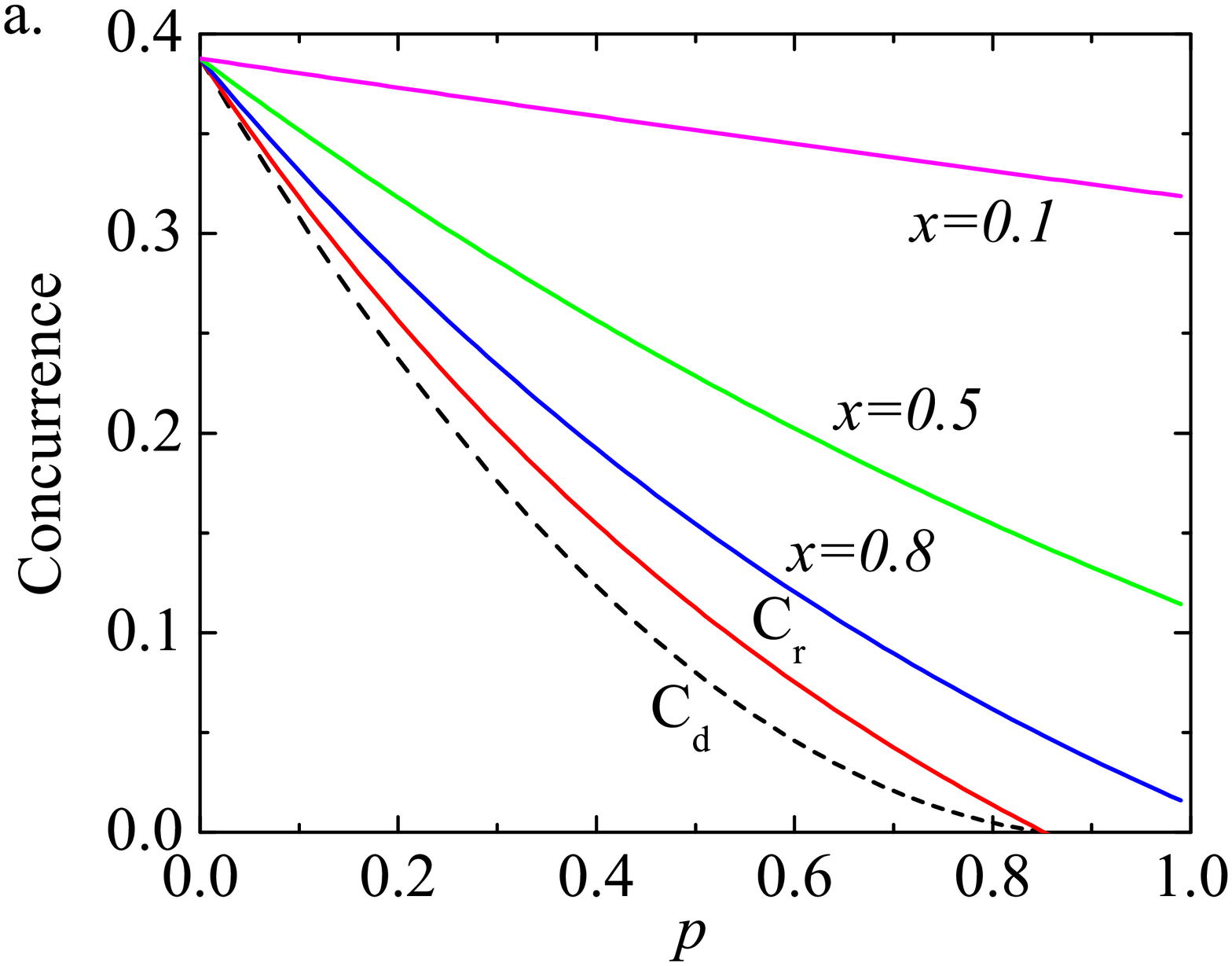}
\includegraphics[width=0.9\columnwidth]{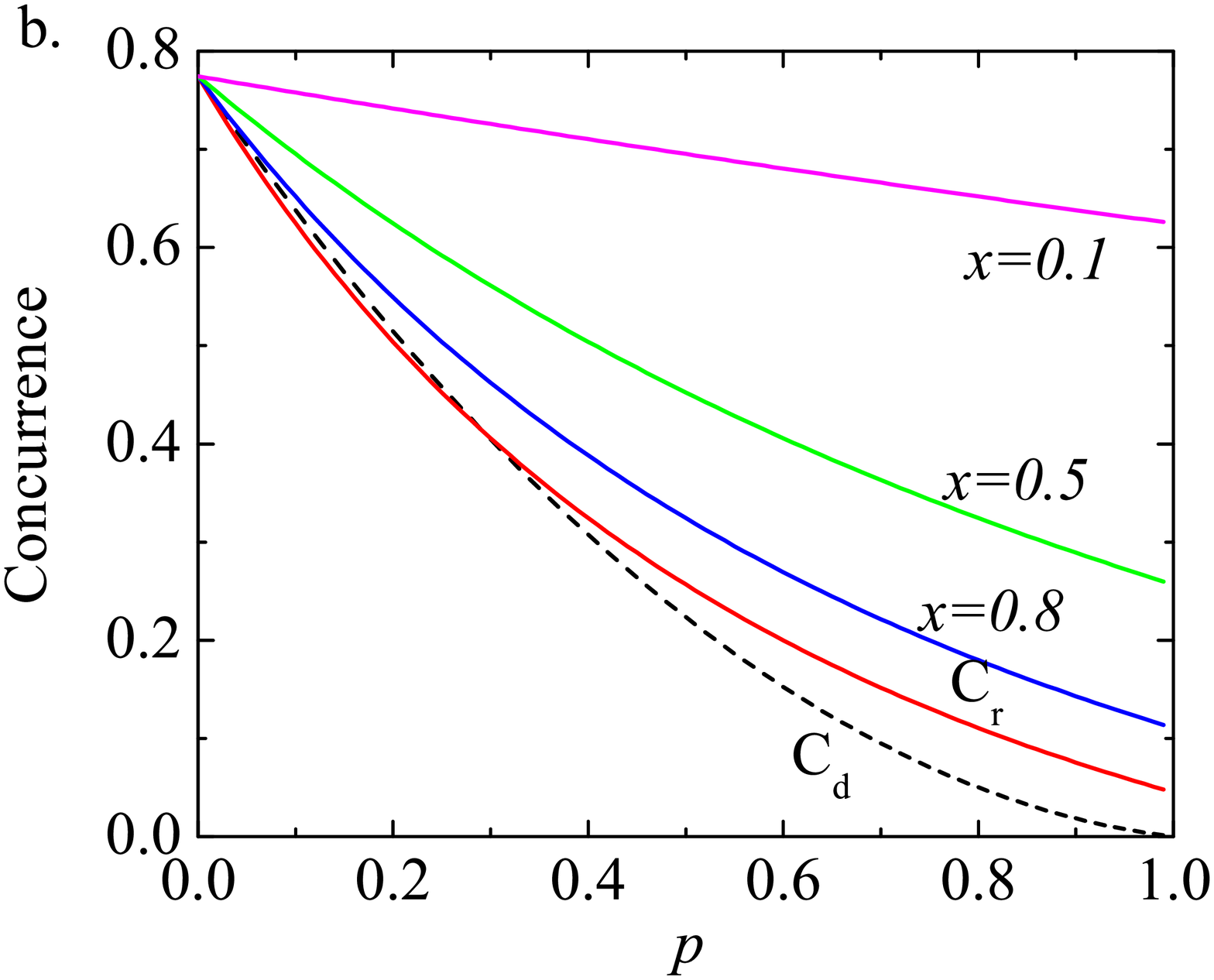}
\caption{(Color online) Concurrence as a function of damping probability $p$. (a) $\alpha=0.7, \beta=0.35, \gamma=0.4, \delta=0.48$. (b) $\alpha=0.10, \beta=0.55, \gamma=-0.60, \delta=0.57$. The dashed lines correspond to the concurrence after amplitude damping, and the red solid lines with $C_{r}$ is the concurrence of the result of the scheme in Sec. IV. The other three curves with $x=0.8, x=0.5$, and $x=0.1$ are the concurrence of the extended scheme in Sec. VII. } 
\end{figure*}

On comparing the concurrences $C_{d}(p)$ and $C_{r}(p)$, we find the following features: 

1) If $|\alpha\delta-\beta\gamma|<|\delta|^{2}$, the entanglement vanishes at $p=|\alpha\delta-\beta\gamma|/\delta ^{2}$ both for the damping state and the recovered state. This is called entanglement sudden death(ESD) \cite{Yu}. Beyond the ESD point, the state of the system is separable and the entanglement can not be recovered in this simple scheme [Fig. 3(a)]. However, we will show in section V that the entanglement of the system beyond the ESD point can also be recovered by simply extending this scheme. 

2) If $|\alpha |\geq  |\delta |$, the final concurrence is always larger than the damping concurrence. Thus in this case the entanglement can always be partially protected [Fig. 3(a)]. 

3) If $\alpha < \delta$, the final concurrence is higher than the damped concurrence only for
\begin{equation}
p>[\sqrt{(1-|\alpha|^{2}+2|\delta|^{2})^{2}-4|\delta|^{2}}-(1-|\alpha|^{2})]/2|\delta|^{2}.
\end{equation}
Otherwise, the concurrence decreases [Fig. 3(b)]. However, in the extended scheme shown in section V we can always choose suitable parameters such that the concurrence always increases for all $p$ values.

If the ancilla qubits are not in $|00\rangle$ state, we can do additional procedures as discussed in the previous section. For example, if the ancilla qubits are in the $|01\rangle$, we add an additional ancilla which is in $|0\rangle$ state. After applying a Hadamard gate on the ancilla qubit, we apply a CNOT gate on the second qubit and the ancilla qubit. If the ancilla qubit is measured to be in $|0\rangle$ state, we find that the density matrix of the system after tracing out the environment is the same as Eq. (13) (see appendix B for calculations). Therefore even if we do not get $|00\rangle$ result, additional procedure can have some probabilities to  protect the quantum entanglement.

\section{V. implementation with linear optics}

\begin{figure}
\includegraphics[width=0.99\columnwidth]{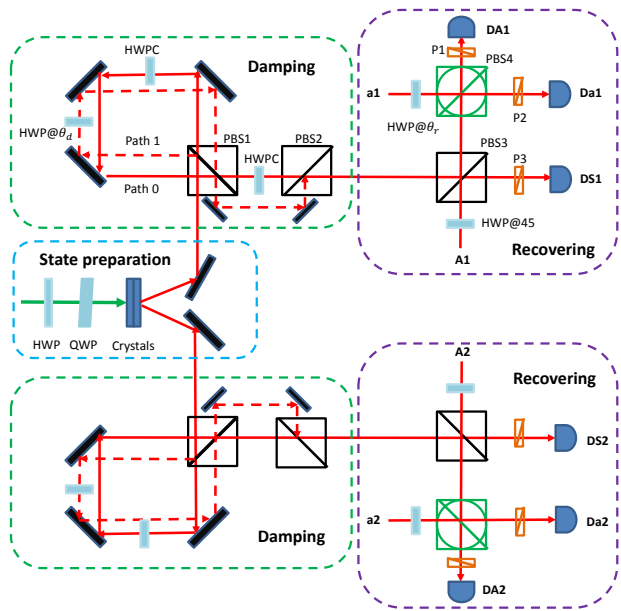}
\caption{(Color online) The experimental scheme for protecting quantum entanglement via Hadamard and CNOT gates. PBS1, PBS2, PBS3 are polarizing beam splitter with horizontal and vertical basis, while PBS4 is polarizing beam splitter with $\pm45^{o}$ basis. P1, P2, P3 are polarizer used for qubit state analysis with P1 and P2 are in the H polarization while P3 can be changed to do the quantum state tomography. }
\end{figure}

A possible experimental scheme with linear optics system is discussed in this section. The experimental setup includes three parts (Fig. 4): entangled state generation, amplitude damping simulation and recovering operations. The setup for entangled state generation and the amplitude damping simulation are the same as that in the weak measurement reversal scheme \cite{Sun2, Kim}.  We briefly describe these two parts. The polarization-entangled photon pair can be generated by two adjacent type-I crystals and the outcome state is given by
\begin{equation}
|\psi\rangle= \alpha |HH\rangle + \beta |VV \rangle,
\end{equation}
where $H$ is the horizontal polarization which is denoted as $|0\rangle$ state, while $V$ is the vertical polarization which is denoted as $|1\rangle$ state. Here $\alpha$ and $\beta$ are two complex number and their values can be controlled by a half-wave-plate (HWP) and a tilted quarter-wave plate (QWP) before the crystals. We can also use a HWP to rotate one of the photons or use a type-II phase matching to generate entangled photons with orthogonal polarizations:
\begin{equation}
|\psi\rangle= \alpha |HV\rangle + \beta |VH \rangle.
\end{equation}

The two entangled qubits are then spatially separated and each qubit goes through a displaced-Sagnac interferometer that simulate the amplitude damping. The $H$ photon travels along the solid line (path 0) while the $V$ photon travels along the dashed line (path 1). There is a HWP with angle $\theta_{d}$ in the path 1 which rotates the $V$ state to a superposition of $H$ state and $V$ state, i.e., $|V\rangle \rightarrow \cos\theta_{d} |V\rangle + \sin\theta_{d} |H\rangle$. This is equivalent to an amplitude damping with $\sqrt{p}=\sin\theta_{d}$. While the HWPC with angle 0 in the path 0 does not rotate the polarization and it is just used to compensate the optical path difference. Then the photon in path 0 and path 1 are combined through PBS2 and the resulting state is given by Eq. (11).  

For the recovering parts, we need to apply two Hadamard gates for the two ancilla qubits (a1, a2) and two CNOT gates. The system qubits are the controlled qubits while the two ancilla qubits (a1, a2) are the target qubits. Initially, both a1 and a2 are in the horizontal polarizations. The Hadamard gates are implemented by HWP with angle $\theta_{r}=\cot^{-1}(\cos\theta_{d})$. After the Hadamard gates the ancilla qubits are in a suppersposition state. Then CNOT gates need to be applied for each pair of system qubit and the ancilla qubit. Several methods for CNOT gates based on linear optics systems have also been discussed \cite{KLM,Okamoto, Pittman2001, Pittman2003}. Here we propose to use the scheme discussed by Pittman et. al. \cite{Pittman2003} where only one auxiliary qubit is needed for each CNOT gate. Initially both auxiliary qubits A1 and A2 are in the horizontal polarization. After passing through a HWP, the states of A1 and A2 become a superposition state $1/\sqrt{2}(|H\rangle +|V\rangle)$. The system qubits and the auxiliary qubits A1 and A2 interfere at the polarizing beam splitter PBS3 where $H$ photon will be transmitted while the $V$ photon will be reflected. Then they interfere with the ancilla qubits (a1, a2) in the PBS4 which is a polarizing beam splitter with basis rotated by $45^{o}$, i.e., $|F\rangle=1/\sqrt{2}(|H\rangle+|V\rangle)$ photon will transmit while $|S\rangle=1/\sqrt{2}(|V\rangle-|H\rangle)$ photon will be reflected. After the PBS4, single photon detectors DA1 and DA2 are used to measure the output states of A1 and A2. A horizontal polarizer P1 is placed in front the detectors, and if we detect one and only one photon with horizontal polarization in both DA1 and DA2, then the probabilistic CNOT gate is successful. The success probability is $1/8$. If at the same time both single photon detectors Da1 and Da2 detect one and only one photon with horizontal polarization ($|00\rangle _{A}$), we keep the result that we get from the detectors DS1 and DS2. We repeat the same procedure simply by changing the polarizations of polarizer P3, and we can carry out the quantum state tomography from which we can calculate the concurrence. We can then compare this concurrence with the damped concurrence. 

\section{VI. Extended scheme}

Here we discuss how we can improve the result by appropriately extending the scheme in Sec. III. The extended scheme is shown in Fig. 5. Before the system qubits undergoing amplitude damping, we apply the same quantum circuit as in the recovery part to prepare the system in a more robust quantum state. If the ancilla qubits are measured to be $|00\rangle$, the preparation is successful, otherwise we should discard the result. The system undergoes amplitude after the preparation stage.  Finally, we do the same recovery procedure as in Sec. IV to restore the quantum state and quantum entanglement.

The state of the system after the successful preparation step can be readily obtained from Eq. (27) in appendix A. The only differences are here $q=1$ and $\theta=\theta_{1}$ where $\theta_{1}$ is the rotation angle of the Hadamard gate in the preparation step. If we define $x=\tan^{2}\theta_{1}$, we obtain
\begin{equation}
|\psi_{P}\rangle=\frac{1}{N_{1}}(\alpha |00\rangle +\beta \sqrt{x} |01\rangle +\gamma \sqrt{x} |10\rangle + \delta x |11\rangle )
\end{equation}
where $N_{1}=\sqrt{|\alpha|^{2}+|\beta|^{2}x+|\gamma|^{2}x+|\delta|^{2}x^{2}}$. The success probability is $N_{1}^{2}/(1+x)^{2}$. If $\theta_{1}$ is chosen such that $x$ is less than $1$, the system uncollapses toward the ground state which has the similar effect as weak measurement \cite{Korotkov, Kim}. The ground state is uncoupled to the environment and is less vulnerable to decoherence. Different from the weak measurement scheme \cite{Korotkov}, here we do not need to wait for the null-result weak measurement.  

\begin{figure*}
\includegraphics[width=1.8\columnwidth]{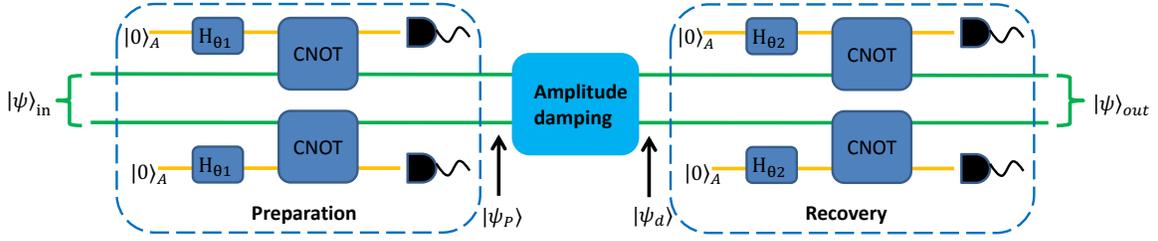}
\caption{(Color online) Improved scheme to protect the quantum entanglement without weak measurement. } 
\end{figure*}

After preparing the system in the state shown in Eq. (18), the system undergoes the amplitude damping and the recovery procedure. In the recovery process, we choose the rotation angle of the Hardamard gate such that $xqy=1$ where $y\equiv \tan ^{2} \theta_{2}$ ($\theta_{2}$ is the rotation angle of the Hardamard gate in the recovery procedure). We measure the state of the ancilla qubits, and if we get $|00\rangle $ result, the density matrix of the system becomes
\begin{widetext}
\begin{equation}
\rho _{f}=\frac{1}{N_{2}} \left( \begin{array}{cccc}
|\alpha|^{2}+p x|\beta|^{2}+p x|\gamma|^{2}+p^{2} x^{2}|\delta|^{2} & \alpha \beta^{*}+p x \gamma\delta^{*} & \alpha \gamma^{*}+p x \beta\delta^{*} &  \alpha \delta^{*} \\
\alpha^{*}\beta + p x \gamma^{*} \delta  & |\beta|^{2}+ p x |\delta|^{2} & \beta \gamma^{*}  & \beta \delta^{*} \\
 \alpha ^{*} \gamma +  p x \beta ^{*}\delta & \beta ^{*} \gamma & |\gamma|^{2}+  p x |\delta|^{2} & \gamma \delta ^{*} \\
\alpha ^{*} \delta & \beta^{*} \delta & \gamma^{*} \delta &  |\delta|^{2} 
\end{array} \right)
\end{equation}
\end{widetext}
where $N_{2}=1+p x(|\beta|^{2}+|\gamma|^{2}+2|\delta|^{2})+p^{2} x^{2}|\delta|^{2}$ is the normalization factor. The concurrence of the final quantum state is given by
\begin{equation}
C_{r}(p,x)=\max \{0,\frac{2(|\alpha \delta -\beta \gamma| - p  x |\delta |)}{1+p x (1-|\alpha|^{2}-|\delta|^{2})+p^{2}x^{2}|\delta|^{2}}\}.
\end{equation}  
On comparing this concurrence with concurrence in Eq. (14) and the damped concurrence $C_{d}(p)$ in Eq. (13), we find several new features. (i) When $x=1$ the new concurrence return back to the result in Eq. (14). This clearly shows that the scheme in Sec. VI is a special case the extended scheme in this section. (ii) For arbitrary initial state if $x$ is small enough $C_{r}(p,x)$ here can be always larger than $C_{d}(p)$ for arbitrary $p$ (Fig. 3a and 3b). This is different from the scheme discussed in the Sec. III where there are always some states that the concurrence can not be improved. (iii) If $x$ is less than $1$, the recovered concurrence can be nonzero even if the damped concurrence is zero (Fig. 3a). This means that the quantum entanglement can also be partially recovered even beyond the ESD point, which never happens in the scheme described in Sec. III. In the limit that $x\rightarrow 0$, the concurrence is $\max \{0,2|\alpha\delta-\beta \gamma|\}$ which is the concurrence of the initial state, i.e., the damped quantum state is recovered back to the initial state. However, we should note that in this case the success probability approaches zero. Therefore there is a tradeoff between the success probability and the entanglement protection. 

\begin{figure*}
\includegraphics[width=0.9\columnwidth]{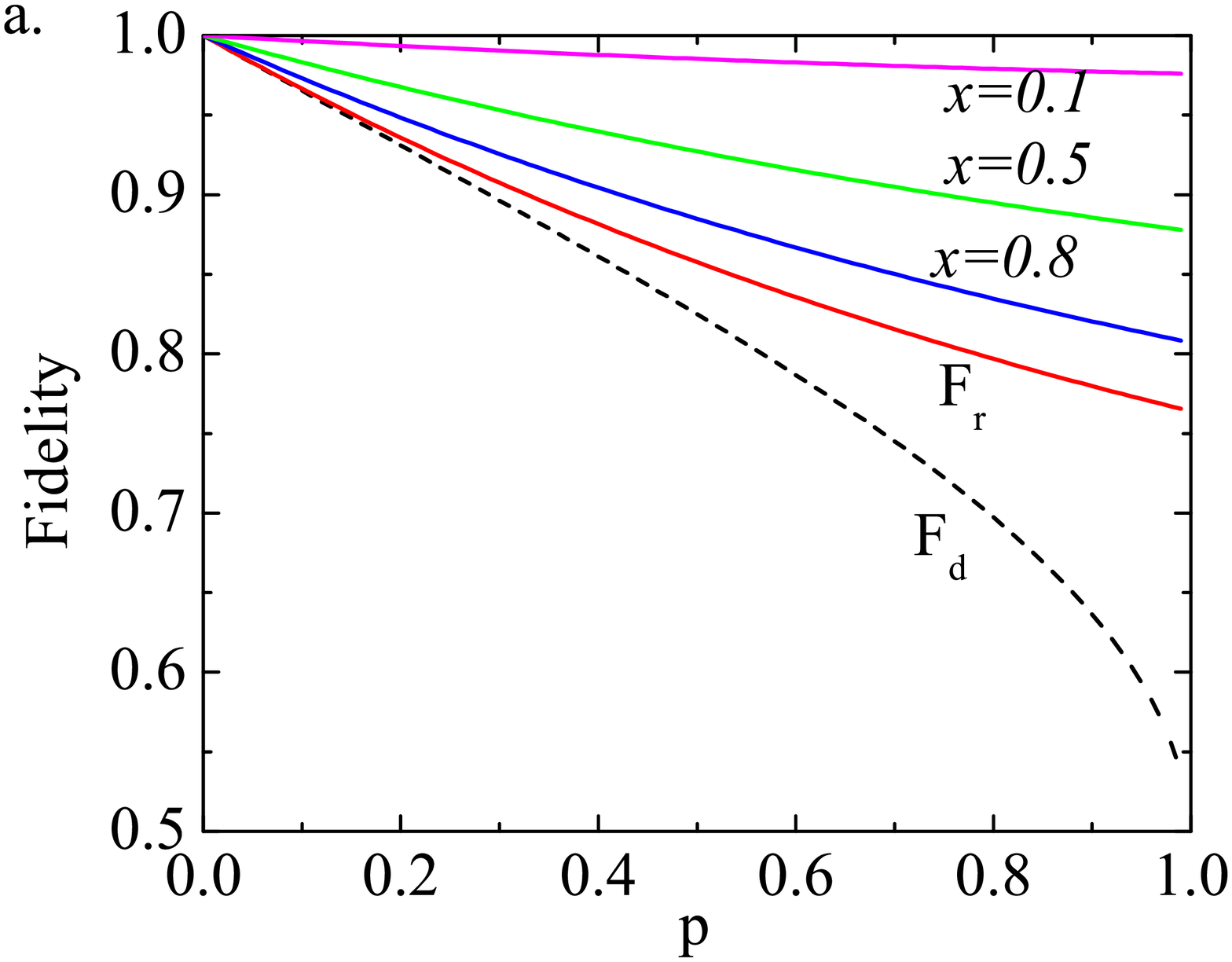}
\includegraphics[width=0.9\columnwidth]{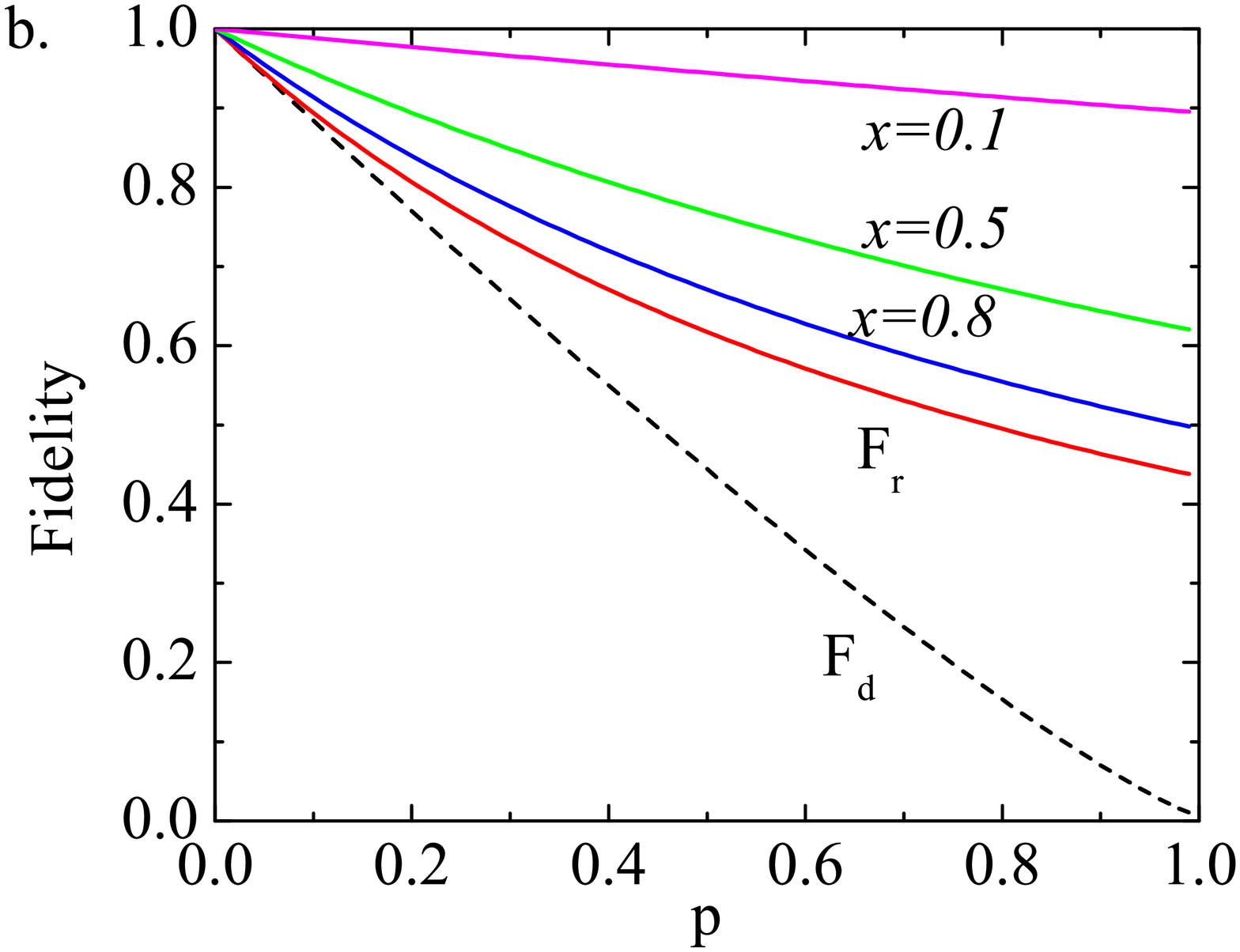}
\caption{(Color online) The fidelity between initial state and the final state as a function of damping probability $p$. (a) $\alpha=0.7, \beta=0.35, \gamma=0.4, \delta=0.48$. (b) $\alpha=0.10, \beta=0.55, \gamma=-0.60, \delta=0.57$. The dashed lines correspond to the fidelity after amplitude damping, and the red solid lines with $F_{r}$ is the fidelity of the result of the scheme in Sec. IV. The other three curves with $x=0.8, x=0.5$, and $x=0.1$ are the fidelity of the extended scheme.
} 
\end{figure*} 

\section{VII. Fidelity}

In the previous sections, we showed that the quantum entanglement can be partially protected in our schemes. Here we show that the quantum state fidelity can also be protected. The fidelity between the two quantum states is given by \cite{Jozsa}
\begin{equation}
F(\rho_{i},\rho_{f})=[Tr(\sqrt{\sqrt{\rho_{i}}\rho_{f}\sqrt{\rho_{i}}})]^{2}=Tr(\rho_{i}\rho_{f})
\end{equation}
where $\rho_{i}$ and $\rho_{f}$ are the initial and final state, respectively. The second identity is valid when the initial state is pure which is the case in our paper. 

For the damped state in Eq. (11), we can calculate the fidelity between this state and the initial state and it is given by 
\begin{eqnarray}
F_{d}(p)&=& (\alpha ^{2}+\sqrt{q}\beta^{2}+\sqrt{q}\gamma ^{2}+q \delta ^{2})^{2} + 4 p \sqrt{q} \alpha \beta  \gamma \delta \nonumber \\ &+& p[\alpha ^{2} + q \delta ^{2}](\beta ^{2}+\gamma ^{2})+p^{2} \alpha ^{2} \delta ^{2}.
\end{eqnarray}
For the recovered quantum state by the scheme in Sec. III (See Eq. (13)), the fidelity is 
\begin{equation}
F_{r}(p)=\frac{1+4p\alpha\beta\gamma\delta+p(\alpha^{2}+\delta^{2})(\beta^{2}+\gamma^{2})+p^{2}\alpha^{2}\delta^{2}}{1+p(1-\alpha ^{2}+\delta^{2})+p^{2}\delta ^{2}}
\end{equation}
For the recovered quantum state by the extended scheme in Sec. V (Eq. (20)), the fidelity is 
\begin{equation}
F_{r}^{'}(p)=\frac{1+4px\alpha\beta\gamma\delta+px^{2}(\alpha^{2}+\delta^{2})(\beta^{2}+\gamma^{2})+p^{2}x^{2}\alpha^{2}\delta^{2}}{1+p x(1-\alpha ^{2}+\delta^{2})+p^{2}x^{2}\delta ^{2}}
\end{equation}
Comparing Eq. (23) and Eq. (24), we find that $F_{r}(p)$ is a special case of $F_{r}^{'}(p)$ when $x=1$. We can also find that when $x\rightarrow 0$, the fidelity $F_{r}^{'}(p)$ of the extended scheme approaches $1$ which means that the quantum state is recovered back to the initial state. Two examples are given in Fig. 6(a) and Fig. 6(b) where we can see that the fidelity can be controlled by the parameter $x$. One can see that smaller $x$ gives higher fidelity. Therefore, the fidelity of the quantum state can also be well preserved in the extended scheme. However we should also notice that the success probability decreases with smaller $x$.

\section{VIII. summary}

In summary, we showed that an arbitrary two-qubit pure state under amplitude damping in a weak measurement can be recovered using Hadamard and CNOT gates. If a two-qubit system is undergoing amplitude damping (but without weak measurement), we showed using similar technique that quantum entanglement can be partially protected.  A proof-of-principle experiment for protecting quantum entanglement in the linear optics system  is also discussed. Using the same quantum circuit diagram we can prepare the system in a more robust state which is less subjected to decoherence. In this case the quantum entanglement can be recovered more efficiently and the quantum state beyond the ESD point can also be partially recovered. We also show that the fidelity of the quantum states can be protected, particularly under the extended scheme.

\section{Acknowledgment}
This work is supported by grants from the King Abdulaziz City for Science and Technology (KACST) and from the Qatar National Research Fund (QNRF) under the NPRP project 4-520-1-083. MA gratefully acknowledge the hospitality at the Institute for Quantum Science and Engineering and Department of Physics and Astronomy, Texas A\&M University, where this work was done.

\begin{widetext}

\section{Appendix A: Quantum state recovery in a weak measurement}

When a two-qubit pure state undergoes amplitude damping and a null-result for the weak measurement, the system evolves to
Eq. (6). We add two ancillas which are initially in the $|00\rangle $ state. First we apply a Hadamard gate with angle $\theta$ for each ancilla, the state of the system becomes
\begin{eqnarray}
|\psi_{1}\rangle &=&\frac{1}{N_{d}}[\alpha |00\rangle _{S}+ \beta \sqrt{q} |01\rangle _{S}+ \gamma \sqrt{q}|10\rangle_{S} +\delta q|11\rangle_{S} ] \otimes (\cos\theta |0\rangle _{A}+\sin\theta |0\rangle _{A})(\cos\theta |0\rangle _{A}+\sin\theta |0\rangle _{A}) \nonumber \\ &=& \frac{\alpha }{N_{d}}|00\rangle _{S} (\cos ^{2}\theta |00\rangle _{A}+\cos \theta \sin \theta |01\rangle _{A}+\cos \theta \sin \theta |10\rangle _{A}+\sin^{2} \theta |11\rangle _{A})+\frac{\beta \sqrt{q} }{N_{d}}|01\rangle _{S} (\cos ^{2}\theta |00\rangle _{A}+\cos \theta \sin \theta |01\rangle _{A} \nonumber \\ &+& \cos \theta \sin \theta |10\rangle _{A}+\sin^{2} \theta |11\rangle _{A})+\frac{\gamma \sqrt{q}}{N_{d}}|10\rangle _{S} (\cos ^{2}\theta |00\rangle _{A}+\cos \theta \sin \theta |01\rangle _{A}+\cos \theta \sin \theta |10\rangle _{A}+\sin^{2} \theta |11\rangle _{A})\nonumber \\ &+& \frac{\delta q}{N_{d}}|11\rangle _{S} (\cos ^{2}\theta |00\rangle _{A}+\cos \theta \sin \theta |01\rangle _{A}+\cos \theta \sin \theta |10\rangle _{A}+\sin^{2} \theta |11\rangle _{A})
\end{eqnarray}
Then two CNOT gates are separately applied to each pair of the system qubit and the ancilla qubit. The system qubits are the controlled qubits while the ancillas are the target qubits. We obtain
\begin{eqnarray}
|\psi_{2}\rangle &=& \frac{\alpha }{N_{d}}|00\rangle _{S} (\cos ^{2}\theta |00\rangle _{A}+\cos \theta \sin \theta |01\rangle _{A}+\cos \theta \sin \theta |10\rangle _{A}+\sin^{2} \theta |11\rangle _{A})+\frac{\beta \sqrt{q} }{N_{d}}|01\rangle _{S} (\cos ^{2}\theta |01\rangle _{A}+\cos \theta \sin \theta |00\rangle _{A} \nonumber \\& &+ \cos \theta \sin \theta |11\rangle _{A}+\sin^{2} \theta |10\rangle _{A})+\frac{\gamma \sqrt{q}}{N_{d}}|10\rangle _{S} (\cos ^{2}\theta |10\rangle _{A}+\cos \theta \sin \theta |11\rangle _{A}+\cos \theta \sin \theta |00\rangle _{A}+\sin^{2} \theta |01\rangle _{A})\nonumber \\ & &+ \frac{\delta q}{N_{d}}|11\rangle _{S} (\cos ^{2}\theta |11\rangle _{A}+\cos \theta \sin \theta |10\rangle _{A}+\cos \theta \sin \theta |01\rangle _{A}+\sin^{2} \theta |00\rangle _{A}) \\
&=& \frac{\cos^{2}\theta} {N_{d}} \{\alpha |00\rangle + \beta \sqrt{q} \tan\theta  |01\rangle + \gamma \sqrt{q} \tan\theta |10\rangle  +\delta q \tan^{2}\theta |11\rangle \}_{S}\otimes |00\rangle _{A} 
+ \frac{\sin\theta \cos\theta }{N_{d}} \{\alpha |00\rangle  +  \beta \sqrt{q} \cot\theta |01\rangle \nonumber \\ & &+ \gamma \sqrt{q}\tan\theta |10\rangle  +\delta q|11\rangle \}_{S}\otimes |01\rangle _{A}  
 + \frac{\sin\theta \cos\theta} {N_{d}}\{\alpha |00\rangle + \beta \sqrt{q} \tan\theta |01\rangle + \gamma \sqrt{q} \cot\theta |10\rangle +\delta q|11\rangle \}_{S}\otimes |10\rangle _{A} \nonumber \\ & &+  \frac{\sin ^{2}\theta }{N_{d}} [\alpha |00\rangle + \sqrt{q} \cot\theta (\beta  |01\rangle + \gamma  |10\rangle ) +\delta q \cot^{2}\theta |11\rangle ]_{S}\otimes |11\rangle _{A} 
\end{eqnarray}
If $\theta$ is chosen to be $\tan^{-1}(1/\sqrt{q})$, the combined state becomes Eq. (9). 

Next we measure the ancilla qubits. From Eq. (9) we see that if we get $|01\rangle$ result the quantum state of the system reduces to
\begin{equation}
|\psi \rangle_{out}=\frac{1}{N_{01}}(\alpha |00\rangle _{S}+ \beta q |01\rangle _{S}+ \gamma |10\rangle _{S}+\delta q|11\rangle _{S})
\end{equation}
where the normalization factor $N_{01}=\sqrt{|\alpha|^{2}+|\beta|^{2} q^{2}+|\gamma|^{2}+|\delta|^{2}q^{2}}$. We add an ancilla qubit which is initially in $|0\rangle$ state. After applying Hadamard gate with angle $\theta1$, the combined state becomes
\begin{equation}
|\psi_{1}^{'}\rangle=\cos\theta(\alpha |00\rangle _{S}+\beta q |01\rangle _{S}+ \gamma |10\rangle _{S}+\delta q|11\rangle _{S})|0\rangle _{A} +\sin\theta(\alpha |00\rangle _{S}+\beta q |01\rangle _{S}+ \gamma |10\rangle _{S}+\delta q|11\rangle _{S})|1\rangle _{A}
\end{equation} 
It follows, on applying CNOT gate on the second qubit and the ancilla, we obtain the state
\begin{equation}
|\psi_{2}^{'}\rangle=\cos\theta(\alpha |00\rangle _{S}|0\rangle _{A}+ \beta q |01\rangle _{S}|1\rangle _{A}+ \gamma |10\rangle _{S}|0\rangle _{A}+\delta q|11\rangle _{S}|1\rangle _{A})+ \sin\theta(\alpha |00\rangle _{S}|1\rangle _{A}+ \beta q |01\rangle _{S}|0\rangle _{A}+ \gamma |10\rangle _{S}|1\rangle _{A}+\delta q|11\rangle _{S}|0\rangle _{A}) 
\end{equation} 
This state can be rewritten as the state in Eq. (9) and we have some probability to recover back the initial state.

Next we calculate the success probability. First the probability of the null result weak measurement for the quantum state in Eq. (5) is given by $P_{0}=N_{d}^{2}$. From Eq. (10), we see that the probability to get the outcomes $|00\rangle, |01\rangle, |10\rangle, |11\rangle $ are
\begin{equation}
P_{00}(q)=\frac{q^{2}}{N_{d}^{2}(1+q)^{2}}, P_{01}(q)=\frac{q }{N_{d}^{2}(1+q)^{2}}, P_{10}(q)=\frac{q }{N_{d}^{2}(1+q)^{2}}, 
P_{11}(q)=\frac{1}{N_{d}^{2}(1+q)^{2}},
\end{equation}
respectively, where we neglect the normalization factors in the medium stages because they can alway be canceled out when we calculate the success probability.

The success probability for only one recovery procedure is given by
\begin{equation}
P_{N=1}=P_{0}P_{00}(q)=\frac{q^{2}}{(1+q)^{2}}.
\end{equation}
If we get $|01\rangle$ or $|10\rangle$ result, we can do additional recovery procedure and from Eq. (9) we can see that the success probability is $P_{0}P_{01}(q)P_{010}(q)$ where $P_{010}(q)=q^{2}/(1+q^{2})$. If we get the outcome $|11\rangle$, we repeat the recovery procedure and the success probability is $P_{0}P_{11}(q)P_{00}(q^{2})$. Thus the success probability for two recovery procedures is given by
\begin{equation}
P_{N=2}=P_{0}P_{00}(q)+2P_{0}P_{01}(q)P_{010}(q)+P_{0}P_{11}(q)P_{00}(q^{2}).
\end{equation} 
Similarly, for three recovery procedures, the success probability is given by
\begin{equation}
P_{N=3}=P_{0}[P_{00}(q)+2P_{01}(q)P_{010}(q)+P_{11}(q)P_{00}(q^{2})
+ 2P_{01}P_{011}(q)P_{010}(q^{2})+2P_{11}P_{01}(q^{2})P_{010}(q^{2})+P_{11}(q)P_{11}(q^{2})P_{00}(q^{4})].
\end{equation} 
where $P_{011}=1/(1+q^{2})$ is the probability that we get the outcome $|1\rangle$ in Eq. (9). 
In principle we can calculate the success probability for any number of repeating procedure. However, in practice we need only about three repeating times because the success probability increases by very small amount for repeating procedures $N>3$.
   
\section{Appendix B: Quantum entanglement protection without a weak measurement}

The quantum state after the amplitude damping is given by Eq. (11). We add two ancilla qubits which are initially in the $|0\rangle$ state. After applying the Hadamard gates and CNOT gates, the state of the system becomes
%\begin{widetext}
\begin{eqnarray}
|\psi_{2}\rangle &=& \cos^{2}\theta \{ \alpha |00\rangle _{S}|00\rangle _{E}+\beta \sqrt{p}|00\rangle _{S}|01\rangle _{E}+\gamma \sqrt{p}|00\rangle _{S}|10\rangle _{E}+\delta p |00\rangle _{S}|11\rangle _{E}+ \beta \sqrt{q} \tan\theta |01\rangle _{S}|00\rangle _{E} +\nonumber \\ & &\delta \sqrt{pq}\tan\theta |01\rangle _{S}|10\rangle _{E} 
+\gamma \sqrt{q} \tan\theta |10\rangle _{S}|00\rangle _{E} +\delta \sqrt{pq}\tan\theta |10\rangle _{S}|01\rangle _{E}+\delta q \tan^{2}\theta|11\rangle _{S}|00\rangle _{E} \} \otimes |00\rangle _{A}  \nonumber \\ &+& \cos\theta \sin\theta \{\alpha |00\rangle _{S}|00\rangle _{E}+\beta \sqrt{p}|00\rangle _{S}|01\rangle _{E} +\gamma \sqrt{p}|00\rangle _{S}|10\rangle _{E}+ \delta p |00\rangle _{S}|11\rangle _{E}+ \beta \sqrt{q}\cot\theta |01\rangle _{S}|00\rangle _{E} + \nonumber \\ & & \delta \sqrt{pq} \cot\theta|01\rangle _{S}|10\rangle _{E}+ \gamma \sqrt{q}\tan\theta |10\rangle _{S}|00\rangle _{E} +\delta \sqrt{pq}\tan\theta |10\rangle _{S}|01\rangle _{E}+\delta q |11\rangle _{S}|00\rangle _{E} \} \otimes |01\rangle _{A}  \nonumber \\
&+& \cos\theta \sin\theta \{\alpha |00\rangle _{S}|00\rangle _{E}+\beta \sqrt{p}|00\rangle _{S}|01\rangle _{E}+\gamma \sqrt{p}|00\rangle _{S}|10\rangle _{E}+\delta p |00\rangle _{S}|11\rangle _{E}+ \beta \sqrt{q}\tan\theta|01\rangle _{S}|00\rangle _{E}+ \nonumber \\ 
& & \delta \sqrt{pq} \tan\theta|01\rangle _{S}|10\rangle _{E} 
+ \gamma \sqrt{q}\cot\theta|10\rangle _{S}|00\rangle _{E} + \delta \sqrt{pq}\cot\theta |10\rangle _{S}|01\rangle _{E}+\delta q |11\rangle _{S}|00\rangle _{E} \}\otimes |10\rangle _{A}  \nonumber \\ &+& \sin^{2}\theta \{\alpha |00\rangle _{S}|00\rangle _{E}+\beta \sqrt{p}|00\rangle _{S}|01\rangle _{E}+\gamma \sqrt{p}|00\rangle _{S}|10\rangle _{E}
+\delta p |00\rangle _{S}|11\rangle _{E}+ \beta \sqrt{q}\cot\theta |01\rangle _{S}|00\rangle _{E}+ \nonumber \\ 
& &\delta \sqrt{pq} \cot\theta |01\rangle _{S}|10\rangle _{E}+ \gamma \sqrt{q}\cot\theta |10\rangle _{S}|00\rangle _{E}  +\delta \sqrt{pq}\cot\theta |10\rangle _{S}|01\rangle _{E}+\delta q \cot^{2}\theta |11\rangle _{S}|00\rangle _{E} \}\otimes |11\rangle _{A} 
\end{eqnarray}
%\end{widetext}

If we measure the ancilla qubits and the outcome is $|00\rangle_{A}$ with $\tan\theta = 1/\sqrt{q}$, we  obtain
%\begin{widetext}
\begin{eqnarray}
|\psi\rangle_{out}&=& \alpha |00\rangle _{S}|00\rangle _{E}+\beta \sqrt{p}|00\rangle _{S}|01\rangle _{E}+\gamma \sqrt{p}|00\rangle _{S}|10\rangle _{E}+\delta p |00\rangle _{S}|11\rangle _{E}+ \beta |01\rangle _{S}|00\rangle _{E} +\delta \sqrt{p}|01\rangle _{S}|10\rangle _{E}+ \nonumber \\ & & \gamma |10\rangle _{S}|00\rangle _{E} 
+\delta \sqrt{p}|10\rangle _{S}|01\rangle _{E}+\delta |11\rangle _{S}|00\rangle _{E} 
\end{eqnarray}
%\end{widetext}
where we neglect the normalization factor.
After tracing out the environment the density matrix of the system is given by Eq. (12). 

If we get other results we can do additional procedure to protect the entanglement. For example if we get $|01\rangle$ result, we add an additional qubit for second qubit. In the recovery procedure, the state evolves as
\begin{eqnarray}
|\psi\rangle_{d}|0\rangle_{A} &=& (\alpha |00\rangle _{S}|00\rangle _{E}+\beta \sqrt{p}|00\rangle _{S}|01\rangle _{E}+\gamma \sqrt{p}|00\rangle _{S}|10\rangle _{E}+\delta p |00\rangle _{S}|11\rangle _{E}+ \beta q |01\rangle _{S}|00\rangle _{E}+\delta \sqrt{p} q|01\rangle _{S}|10\rangle _{E} \nonumber \\& &+ \gamma |10\rangle _{S}|00\rangle _{E} 
+ \delta \sqrt{(p}|10\rangle _{S}|01\rangle _{E}+\delta q |11\rangle _{S}|00\rangle _{E} )\otimes |0\rangle _{A} \\
\stackrel{H_{\theta}}{\longrightarrow}&&(\alpha |00\rangle _{S}|00\rangle _{E}+\beta \sqrt{p}|00\rangle _{S}|01\rangle _{E}+\gamma \sqrt{p}|00\rangle _{S}|10\rangle _{E}+\delta p |00\rangle _{S}|11\rangle _{E}+ \beta q |01\rangle _{S}|00\rangle _{E}+\delta \sqrt{p} q|01\rangle _{S}|10\rangle _{E}\nonumber \\ & &+ \gamma |10\rangle _{S}|00\rangle _{E} 
+\delta \sqrt{(p}|10\rangle _{S}|01\rangle _{E}+\delta q |11\rangle _{S}|00\rangle _{E} )\otimes (\cos\theta|0\rangle _{A}+\sin\theta |1\rangle ) \\
\stackrel{CNOT}{\longrightarrow }&&\sqrt{\frac{q}{1+q}}(\alpha |00\rangle _{S}|00\rangle _{E}+\beta \sqrt{p}|00\rangle _{S}|01\rangle _{E}+\gamma \sqrt{p}|00\rangle _{S}|10\rangle _{E}+\delta p |00\rangle _{S}|11\rangle _{E}+ \beta |01\rangle _{S}|00\rangle _{E}+\delta \sqrt{p}|01\rangle _{S}|10\rangle _{E} \nonumber \\& & + \gamma |10\rangle _{S}|00\rangle _{E} 
+\delta \sqrt{p}|10\rangle _{S}|01\rangle _{E}+\delta |11\rangle _{S}|00\rangle _{E} )\otimes |0\rangle _{A} 
+ \sqrt{\frac{q}{1+q}}(\alpha |00\rangle _{S}|00\rangle _{E}+\beta \sqrt{p}|00\rangle _{S}|01\rangle _{E}\nonumber \\& &+\gamma \sqrt{p}|00\rangle _{S}|10\rangle _{E}  +\delta p |00\rangle _{S}|11\rangle _{E}+ \beta q^{2}|01\rangle _{S}|00\rangle _{E}+\delta \sqrt{p} q^{2}|01\rangle _{S}|10\rangle _{E}+ \gamma |10\rangle _{S}|00\rangle _{E} 
+\delta \sqrt{p}|10\rangle _{S}|01\rangle _{E}\nonumber \\& &+\delta q^{2}|11\rangle _{S}|00\rangle _{E} ) \otimes |1\rangle _{A} 
\end{eqnarray}
where we have chosen $\tan\theta=1/q$ in the last equation. We can see that if we measure $|0\rangle$ for the ancilla qubit the quantum state of the system returns back to the result given by Eq. (36). If we trace out the environment the quantum state of the system is given by Eq. (13). If we measure $|1\rangle$ for the ancilla qubit we just repeat the same procedure with $\tan\theta=1/q^{2}$.

\section{Appendix C: Extended scheme}

On comparing Eq. (5) and  Eq. (17), we note the following changes: $\beta\rightarrow \beta \sqrt{x}$, $\gamma\rightarrow \gamma \sqrt{x}$, and $\delta\rightarrow \delta x$. Therefor, using the same recovery procedure we get the combined quantum state similar to Eqs. (28)-(31) but with $\beta$ replaced by $\beta \sqrt{x}$, $\gamma$ by $\gamma \sqrt{x}$, and $\delta$ by $\delta x$. If our measurement outcome is $|00\rangle$, we obtain
%\begin{widetext}
\begin{eqnarray}
|\psi\rangle_{out}&=& \alpha |00\rangle _{S}|00\rangle _{E}+\beta \sqrt{px}|00\rangle _{S}|01\rangle _{E}+\gamma \sqrt{px}|00\rangle _{S}|10\rangle _{E}+\delta px |00\rangle _{S}|11\rangle _{E}+ \beta \sqrt{qx} \tan\theta |01\rangle _{S}|00\rangle _{E} \nonumber \\ & & +\delta \sqrt{pq}x\tan\theta |01\rangle _{S}|10\rangle _{E}+ \gamma \sqrt{qx} \tan\theta |10\rangle _{S}|00\rangle _{E} 
+\delta \sqrt{pq}x\tan\theta |10\rangle _{S}|01\rangle _{E}+\delta qx \tan^{2}\theta|11\rangle _{S}|00\rangle _{E} 
\end{eqnarray}
%\end{widetext}
If we choose $\theta$ such that $xqy=1$ where $y=\tan^{2}\theta$, we have 
%\begin{widetext}
\begin{eqnarray}
|\psi\rangle_{out}&=&\alpha |00\rangle _{S}|00\rangle _{E}+\beta \sqrt{px}|00\rangle _{S}|01\rangle _{E}+\gamma \sqrt{px}|00\rangle _{S}|10\rangle _{E}+\delta px |00\rangle _{S}|11\rangle _{E}+ \beta  |01\rangle _{S}|00\rangle _{E}+\delta \sqrt{px}|01\rangle _{S}|10\rangle _{E}\nonumber \\ & & + \gamma  |10\rangle _{S}|00\rangle _{E} 
+\delta \sqrt{px}|10\rangle _{S}|01\rangle _{E}+\delta |11\rangle _{S}|00\rangle _{E} 
\end{eqnarray}
%\end{widetext}
On comparing this quantum state with Eq. (36), we can see that the only difference is that $p$ in Eq. (36)
is replaced by $px$ here. Therefore, after tracing out the environment the density matrix in Eq. (19) is different from Eq. (13) by replacing $p$ with $px$. We can also notice similar changes for the concurrences (Eqs. (14) and (20)) and fidelities (Eqs. (23) and (24)).

\end{widetext}

\end{document}